\documentclass[aip,rsi,amsmath,amssymb,reprint]{revtex4-1}
\usepackage{graphicx}
\usepackage[english]{babel}
\usepackage{amsmath,amssymb,graphicx,color}

\begin{document}

\preprint{AIP/123-QED}

\title[]{Smart Table Based on Metasurface for Wireless Power Transfer}

\author{Mingzhao Song}

\affiliation{Department of Nanophotonics and Metamaterials, ITMO University, 197101 Saint Petersburg, Russia}
%
\author{Kseniia Baryshnikova}
\affiliation{Department of Nanophotonics and Metamaterials, ITMO University, 197101 Saint Petersburg, Russia}

\author{Aleksandr Markvart}
\affiliation{Department of Nanophotonics and Metamaterials, ITMO University, 197101 Saint Petersburg, Russia}

\author{Pavel Belov}
\affiliation{Department of Nanophotonics and Metamaterials, ITMO University, 197101 Saint Petersburg, Russia}

\author{Elizaveta Nenasheva}
\affiliation{Giricond Research Institute, Ceramics Co., Ltd., Saint Petersburg 194223, Russia}

\author{Constantin Simovski}
\affiliation{Department of Nanophotonics and Metamaterials, ITMO University, 197101 Saint Petersburg, Russia}
\affiliation{School of Electrical Engineering, Department of Electronics and Nanoengineering, Aalto University, P.O. Box 15500, 00076 Aalto, Finland}

\author{Polina Kapitanova}
\affiliation{Department of Nanophotonics and Metamaterials, ITMO University, 197101 Saint Petersburg, Russia}

\date{\today}

\begin{abstract}
Metasurfaces have been investigated and its numerous exotic functionalities and the potentials to arbitrarily control of the electromagnetic fields have been extensively explored. However, only limited types of metasurface have finally entered into real products. Here, we introduce a concept of a metasurface-based smart table for wirelessly charging portable devices and report its first prototype. The proposed metasurface can efficiently transform evanescent fields into propagating waves which significantly improves the near field coupling to charge a receiving device arbitrarily placed on its surface wirelessly through magnetic resonance coupling. In this way, power transfer efficiency of 80$\%$ is experimentally obtained when the receiver is placed at any distances from the transmitter. The proposed concept enables a variety of important applications in the fields of consumer electronics, electric automobiles, implanted medical devices, etc. The further developed metasurface-based smart table may serve as an ultimate 2-dimensional platform and support charging multiple receivers. 

\end{abstract}

\maketitle

\section{Introduction}
Arbitrary tailor, mold and manipulation of electromagnetic fields is the ultimate goal in electromagnetic research from radio-frequencies to optics. Recently,  metasurfaces have attracted great interests due to its potentials to provide a profound control over electromagnetic fields.~\cite{glybovski2016Metasurface,chen2016Review,Ding2018Gradient} A variety of functionalities are demonstrated for a broad frequency band ranging from microwave to visible light. Initially, the high impedance surfaces were proposed in radio frequency and microwave regimes to reduce the antenna profile and improve the radiation patterns~\cite{Sievenpiper1999High}. Recently these 2D artificial structures were dubbed as metasurfaces~\cite{glybovski2016Metasurface} and their applications have soon spread to more specific areas such as enhancement of magnetic resonance imaging~\cite{slobozhanyuk2016enhancement}. In optical regime, metasurfaces have paved the way for flat optics and photonics~\cite{yu2014flat, kildishev2013planar}. They can be designed to possess the required properties to replace bulky optical components. Different realizations are demonstrated for specific purposes, for instance, beam focusing lens ~\cite{Khorasaninejad2016metalenses}, tunable lens~\cite{arbabi2018mems}, perfect absorbers~\cite{ra2015thin}, wavefront shapers~\cite{yu2011light}, polarizers~\cite{zhao2012twisted} etc. However, the applications of metasurfaces for wireless power transfer (WPT) are yet to be explored.

With an increasing demands for conveniently charging electronic devices, WPT is considered as a promising solution. Much effort has been made to develop different types of WPT systems,~\cite{Hui2014Critical,lu2017review,song2017wireless,assawaworrarit2017robust,krasnok2018coherently,song2016wireless} among which the magnetic resonant coupling has become a hot research topic since it was proposed for the first time in 2007 especially due to its potentials for safe and mid-range charging.~\cite{costanzo2014electromagnetic,kurs2007wireless,karalis2008efficient}  In such systems two or more resonators with the same resonance frequency are coupled by magnetic evanescent fields. The typical charging distance is 2-10 times of the resonator characteristic dimension. The charging distance is limited primarily due to the physical principle of evanescent field decay rather than engineering restrictions. Thus there is little room to further extend the operation distance of a magnetic resonant WPT system in 3D space. One of the ways is to increase the coupling coefficient between two distant resonators with the use of metamaterials. Different metamaterials designs for WPT have been recently introduced and intensively studied.~\cite{urzhumov2011metamaterial,chen2012extremely,dong2017enhanced,wang2011experiments,ho2015planar}
Usually, metamaterials operate as a super-lens focusing near magnetic fields of a transmitter and redirecting it to a receiver.~\cite{wang2011experiments} It helps to improve the WPT efficiency up to 30\% at the fixed distance of the transmitter and receiver placed at the optimized position (or focal points) from two opposite sides of the metamaterial.~\cite{wang2011experiments} However, the whole WPT system with bulk metamaterial is very cumbersome which reduces its chances for practical applications.
\begin{figure*}
\includegraphics[width=\textwidth]{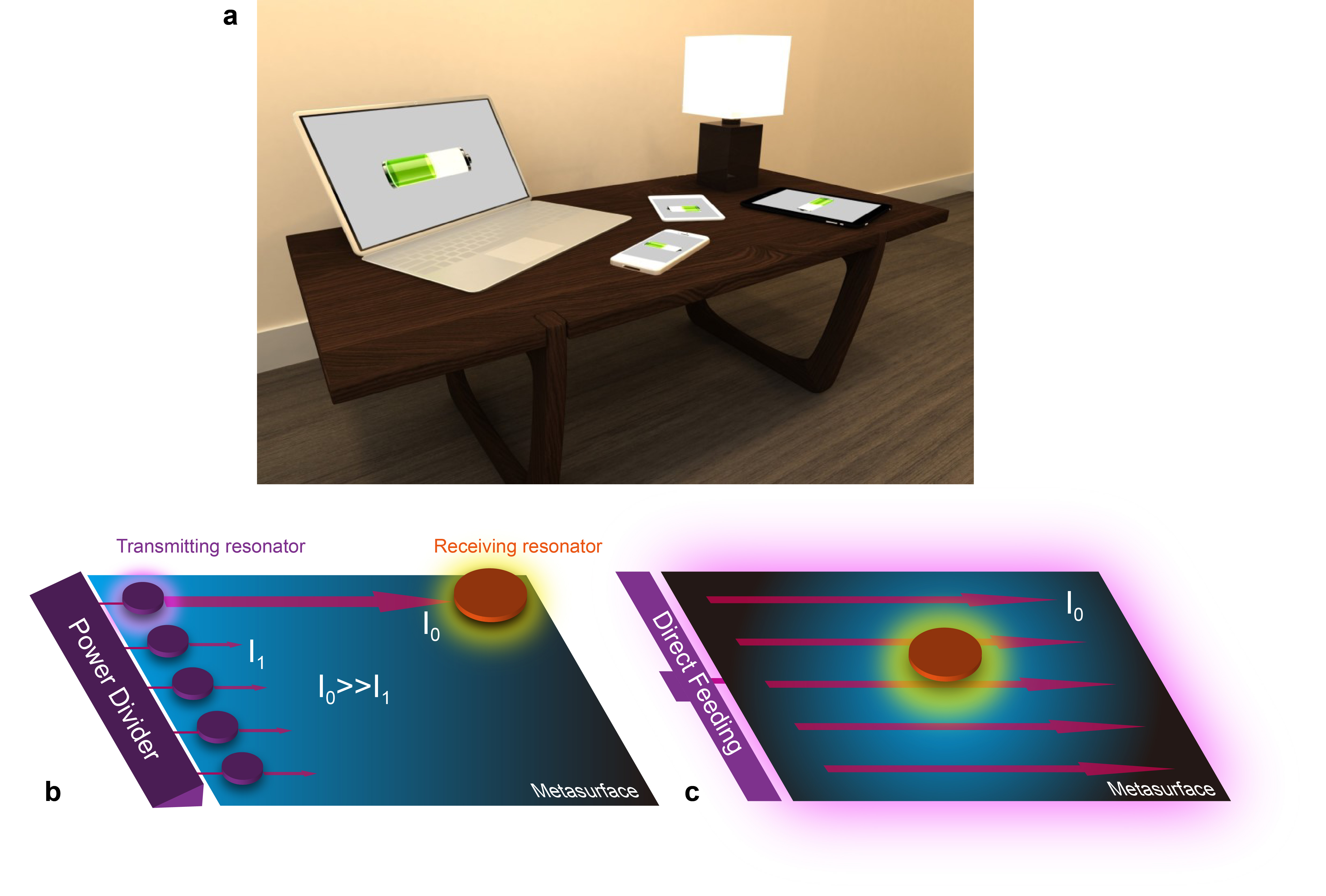}
\caption{(a) Artist's view of multiple electronic devices being powered by a wireless charging table. (b) Metasurface as an intermediary for enhancing WPT performance. (c) Metasurface as a transmitting resonator of WPT system.}
\label{fig1}
\end{figure*}

Unlike the bulk metamaterials, thin planar metasurfaces possess low intrinsic losses and at the same time provide a desired manipulation of the electromagnetic fields. The goal of this paper is to introduce a novel concept of a metasurface-based smart table for WPT applications. In our vision the smart table with the metasurface embedded or placed underneath the desktop allows powering multiple devices simultaneously, regardless of how they are located on the table and oriented relative to each other (see Fig.~\ref{fig1}(a)). In terms of the operational principle we consider different scenarios of the metasurface designs. First, the metasurface plays a role of an intermediary between the transmitter and the receiver enhancing the WPT performances, as shown in Fig.~\ref{fig1}(b). The metasurface comprises multiple power transfer channels. When a receiving resonator is placed on the metasurface and detected, the channel underneath is switched on by activating the corresponding transmitter driven by the power divider. At the same time, the inactivated channel keeps off. Thus the major part of the power will flow only through the activated channel in the form of current $I_0$, whereas the leakage current in other channels is negligible ($I_0 >> I_1$). The advantage of this type of design is that the power can be fully used with minimum losses when the power transfer channel is well designed. But it requires load detection elements which increases the engineering complexity. In the other design, the metasurface itself behaves as a transmitting resonator providing a desired magnetic field profile, for instance, a uniform field distribution or hot spots, as shown in Fig.~\ref{fig1}(c). In this case, the metasurface operates on a certain mode which requires a simple excitation. However, part of the energy will be dissipated in the form of radiation.
Here we report the first scenario and demonstrate the metasurface design which serves as a substrate to transmitting and receiving resonators allowing an efficient WPT up to the distances exceeding the resonator size by an order of magnitude. To demonstrate the function of the metasurface and its possibilities for long-range power transfer, the WPT efficiency from one resonator to another was investigated both numerically and experimentally. The calculations of specific absorption rate (SAR) which is a critical factor in terms of the safety issues was performed. Our design being affordable and compact paves the way to the realization of ubiquitous wireless charging.

\section{Design of WPT system}
Recently, wire media, arrays of parallel metal wires are studied in microwave, terahertz and optical frequency regimes revealing unique electromagnetic properties.~\cite{simovski2012wire} One of these properties is the efficient conversion of evanescent waves into waves propagating in the wire medium.~\citep{rahman2009importance,rahman2010periscope,belov2010experimental} This property is a key prerequisite for the subwavelength imaging in wire-medium endoscopes~\cite{radu2009toward,simovski2012wire,tuniz2013metamaterial,slobozhanyuk2014endoscope}. And it is also the prerequisite of an efficient WPT between two resonators. 

\begin{figure*}
\includegraphics[width=.8\textwidth]{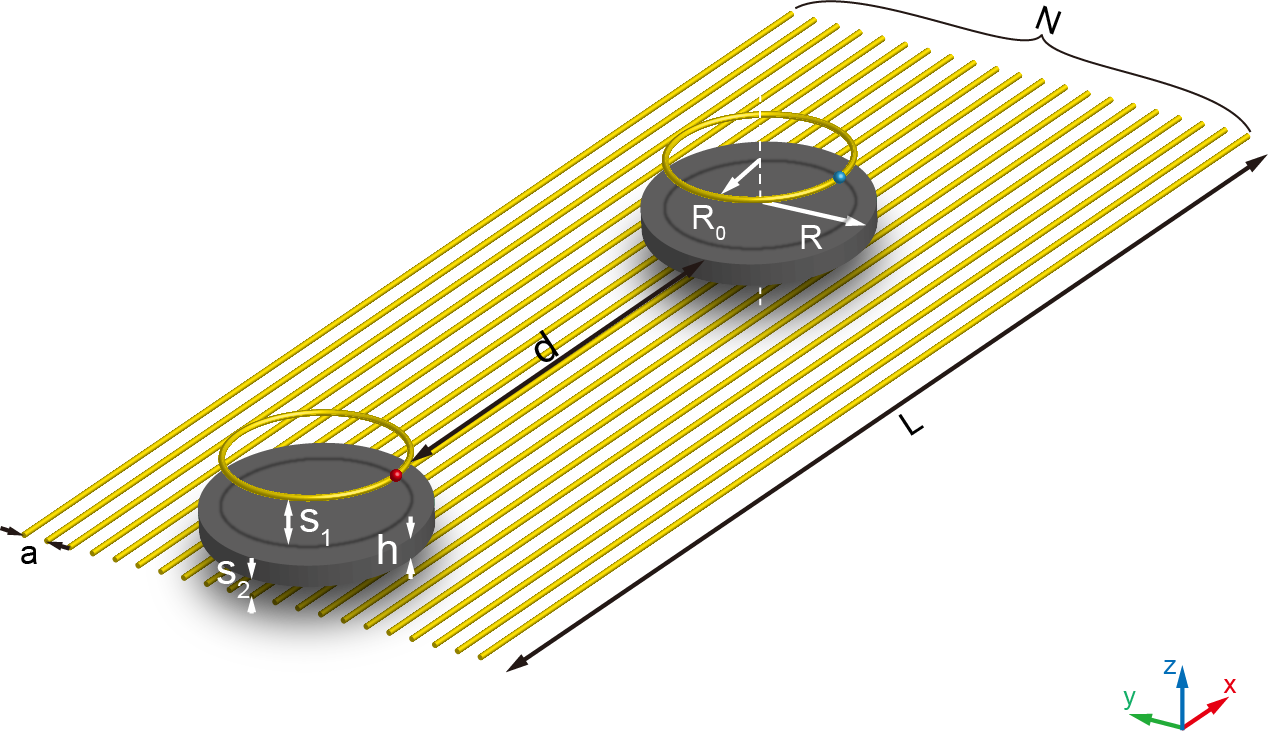}
\caption{Geometry of proposed WPT system.  }
\label{fig2}
\end{figure*}

In our case, the resonant coupling is the coupling of near fields which are packages of evanescent waves. For a disk resonator with colossal permittivity operating at a magnetic dipole mode the electric field is mostly concentrated inside while the magnetic field is maximal near the surface of the disk.~\cite{song2016wireless1} The evanescent tails of the magnetic field offer the WPT with 50$\%$ efficiency at the distances twofold of the resonator diameter.~\cite{song2016wireless1} The further distance increase results in an efficiency decay. However, if a wire medium is located at a distance smaller that the resonator size the resonator may be strongly coupled to the wire medium. Its near magnetic field is converted into propagating modes of the wire medium and the power can be transported from one resonator to another along the wires.  Such mechanism still belongs to WPT because two resonators are not electrically connected. In the present work, we suggest a more practical way than the use of bulk wire medium. A metasurface of parallel wires, a planar analogue of the wire medium, is also capable to convert the evanescent waves into propagating modes and vice versa. It can nicely serve as an intermediary for distant coupling between two disk resonators.

The WPT system under study consisted of a transmitter and a receiver placed above the metasurface as shown in Fig.\ref{fig2}. The transmitter comprised a dielectric disk resonator and a transmitting loop on top of it at the distance $s_1$. 
The dielectric disk had a diameter of $D$ and a thickness of $h$. The loop had a diameter of $D_0$ and was made of a conducting wire with diameter of $w$. The receiver consisted of an identical dielectric resonator and a receiving loop. The receiver was located at a varying distance $d$ from the transmitter. Both transmitter and receiver were placed at height $s_2$ above the metasurface. The last one was a regular array of parallel copper wires with period $a$ satisfying the criterion $w\ll a\ll\lambda$, where $\lambda$ was the wavelength in free space. Thus this parameter was chosen as $a$=10 mm. There was no limitation for the total number of wires in the metasurface, but it depended on the practical needs. Here we used 21 wires to cover 20 cm interval which was the width of our metasurface. The length of the metasurface $L$ was also selected from practical requirements. Two cases of $L$=65 cm and $L$=120 cm were studied.

\section{Results and Discussion}

\subsection{Operational modes of WPT system}

Eigenmode analysis of a single dielectric resonator was conducted, which revealed that all the magnetic modes were identified at the frequencies above 230 MHz, i.e.magnetic dipole (MD), quadrupole (MQ) and octupole (MO) mode hold at 232 MHz, 296 MHz and 347 MHz, respectively, whereas all the electric modes were observed at the frequencies higher than 720 MHz, i.e. electric dipole (ED) and quadrupole (EQ) mode is at 725 MHz and 730 MHz, respectively. Such a wide range of magnetic and electric eigenmodes separation is resulted from the geometric parameters of the dielectric disk resonator. Therefore we mainly target to use the MD mode of the disk which can be excited by a current loop~\cite{song2016wireless1} being aware of the electric type of modes located far away from magnetic types. 

\begin{figure}[b]
 \includegraphics[width=\columnwidth]{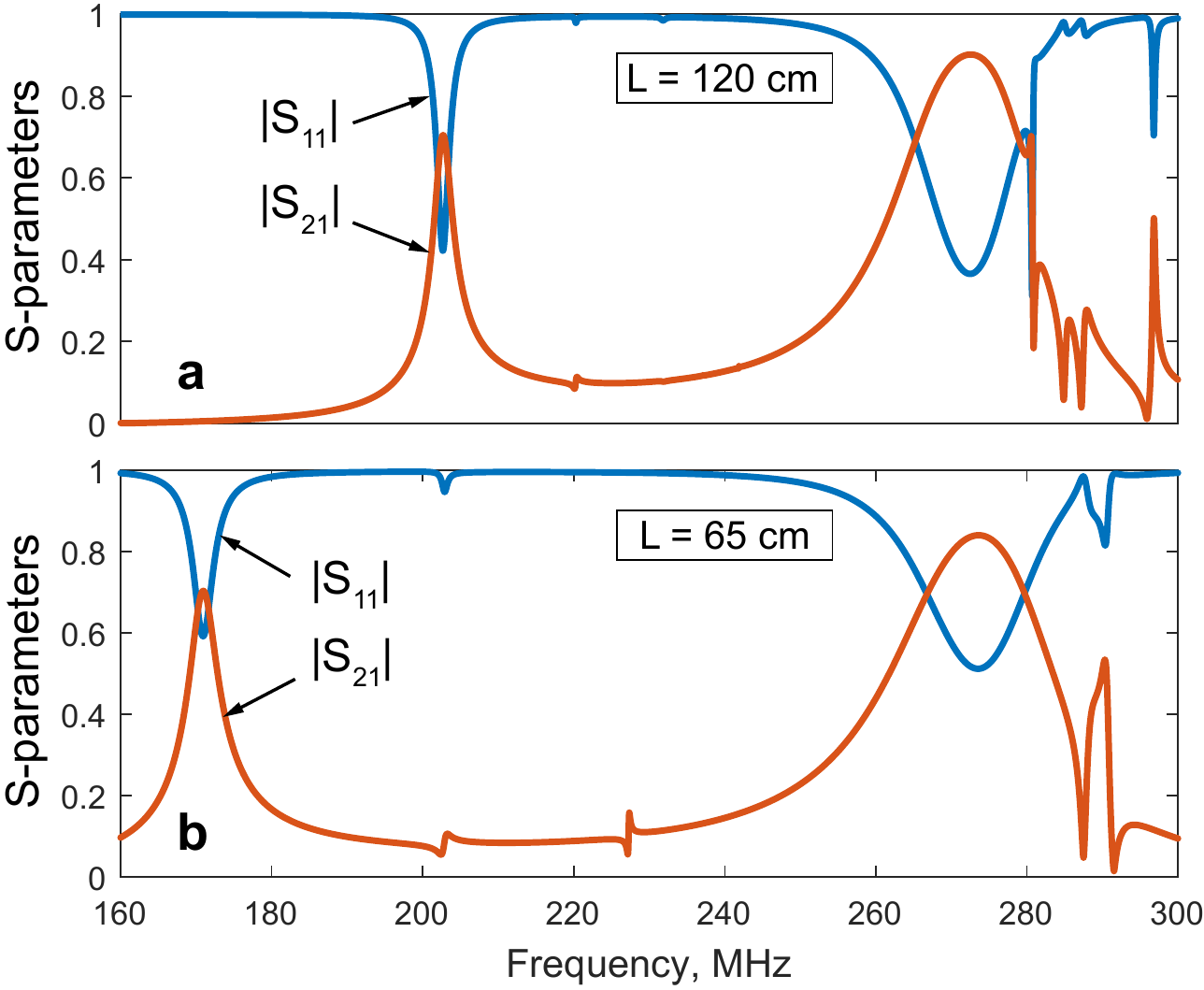}
\caption{Comparison between the simulated S-parameter spectra of the structures with the length of metasurface (a) $L=120$ cm and (b) $L=65$ cm.}
\label{fig3}
\end{figure}

In the proposed WPT system the transmitting resonator was coupled to the metasurface and, via the metasurface, to another disk. Here the frequencies of all modes were shifted with respect to those of an individual disk and could hybridize if the coupling with the metasurface was too strong. For two cases of $L = 65$ cm and $L = 120$ cm, we performed numerical simulations of the whole WPT system using the frequency solver in CST Microwave Studio Suite varying $d$ and keeping the same value $s_2$ which determined the coupling with the metasurface, see Fig.\ref{fig3} for $d$ = 25 cm. For the WPT system with $L$=120 cm in Fig.\ref{fig3} one can see two maxima in the transmission coefficient spectrum -- at frequency nearly equal to 200 MHz and at nearly 270 MHz. For the design with $L$ = 65 cm two peaks are also observed. The WPT efficiency was estimated from the S-parameters using the equation $\eta=|S_{21}|^2$. It was found that for $L=120$ cm the efficiencies of $\eta$=49\% at 200 MHz and $\eta$=81\% at 270 MHz are reached. In the case of metasurface length of $L=65$ cm the efficiencies of $\eta$=49\% at 170 MHz and $\eta$=70\% at 270 MHz are obtained. Thus, for given metasurfaces we found two frequencies for an efficient power transfer for the given distance between resonators.
However, the first transmission resonance shifted to 170 MHz, more than by 15\% compared to the case of $L$=120 cm, whereas the second peak of transmission remained at the same frequency. Thus, this resonance was not a dimensional one of the metasurface and was not affected by its finite size. The main question then arose: which of these two transmission peaks corresponded to the MD mode? 

\begin{figure}
\includegraphics[width=\columnwidth]{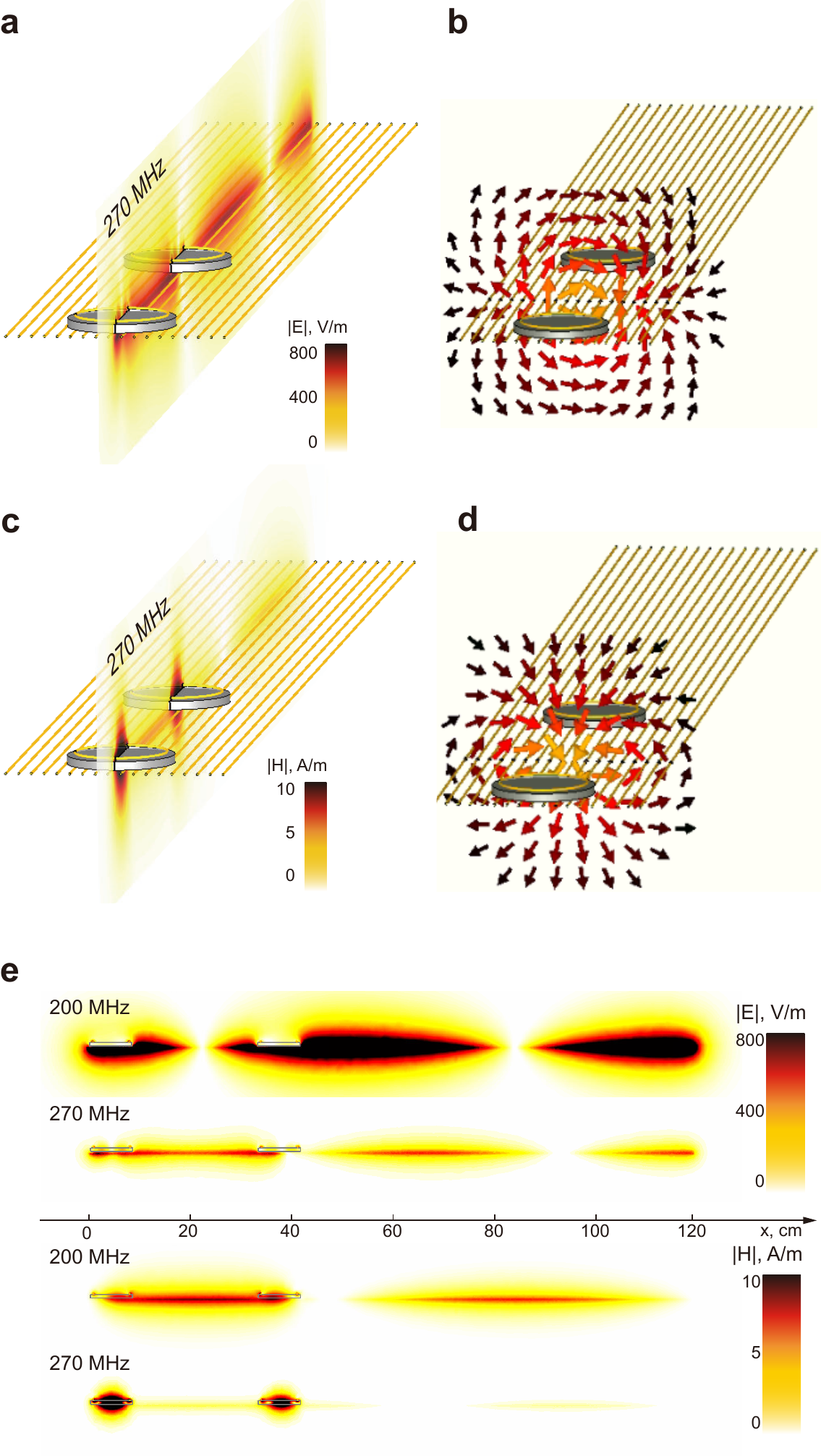}
\caption{ Simulated electric field at 270 MHz of (a) absolute value on the cross section cutting through two disk resonators and (b) vector form on the cross section half way between two resonators. Simulated magnetic field at 270 MHz of (c) absolute value and (d) vector form. (d) Comparison between electric and magnetic fields at 200 MHz and 270 MHz.}
\label{fig4}
\end{figure}

To clarify the nature of both resonances we analyzed the electromagnetic field distributions for the case $L=120$ cm. The electric and magnetic field distributions across the metasurface and across the disk at two frequencies are shown in Fig.\ref{fig4}. As one can see in Fig.\ref{fig4}(e) that at 200 MHz the electric field concentrates at the edges and in the middle of the metasurface forming a typical standing wave pattern.
The symmetry of the standing wave was broken due to the asymmetric placement of the disks. The magnetic field distribution confirmed that the pattern of the electromagnetic field corresponded to the standing wave at 200 MHz where a resonance of the disk and that of a metasurface overlapped. 
It was clear that the coupling at this frequency must be sensitive to the distance $d$. At 270 MHz the electric field kept uniform between the resonators. The minima of $|\bf E|$ at the disks axes was a hint that 270 MHz was the disk resonance of the magnetic type. 
Also, the maximal amplitude of the electric field at 270 MHz was lower than that at 200 MHz, which was good from safety reasons and was another advantage of this operation frequency.
At 270 MHz the features of the standing wave regime were also present in the electromagnetic field distribution but are very weak, and the role of the disk resonance was more important. 
And in the color map of the magnetic field at 270 MHz the axial field concentration was observed which was a typical feature of MD resonance. In the steady regime the reactive part of the mode energy was stored in the disk and did not play any role, whereas the active part -- with a nonzero Poynting vector -- coupled to the metasurface. Fig.~\ref{fig4}(b) and (d) demonstrate that this active power excited a TEM mode in the metasurface and transmitted along it as in a multi-wire transmission line. At a distance $d$ this guided mode excited the same MD mode in the receiving disk that was similarly coupled to the metasurface. Finally, the active energy of the transmitting disk was received by the wire loop applied to the receiving disk resonator.

\begin{figure}[b]
\centering
\includegraphics[width=\columnwidth]{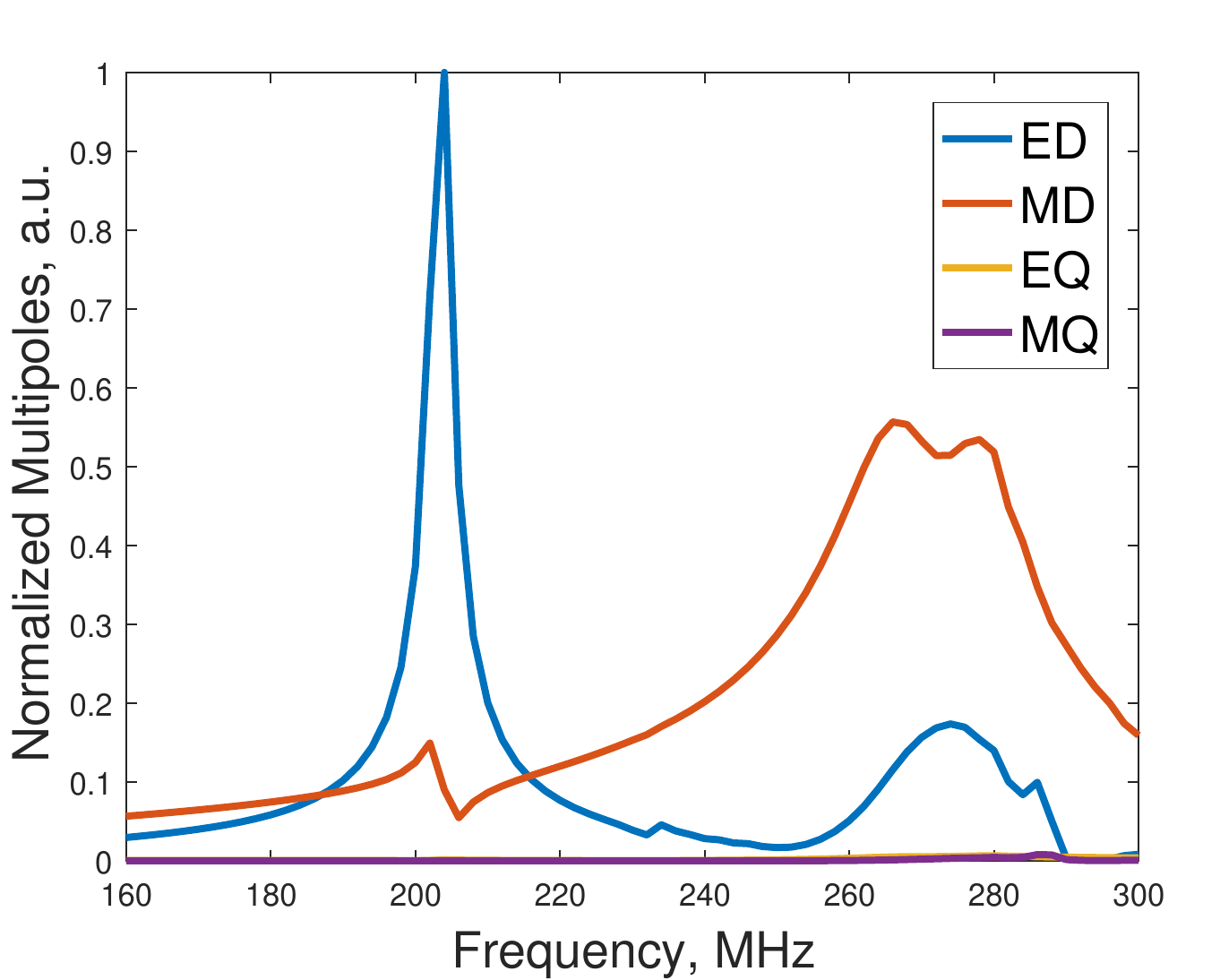}
\caption{Multipole decomposition of the transmitting disk resonator.}
\label{fig5}
\end{figure}

To finally verify the mode type we performed the multipole decomposition of the induced displacement currents inside the transmitting resonator of the WPT system. 
Four terms were taken into consideration, namely, electric dipole (ED), magnetic dipole (MD), electric quadrupole (EQ) and magnetic quadrupole (MQ) modes.~\cite{evlyukhin2016optical,terekhov2017multipolar} 
The calculated multipoles are shown in Fig.\ref{fig5}. A strong ED mode at 200 MHz frequency and the MD mode at 270 MHz were observed, whereas both electric and magnetic quadrupole modes do not resonate below 300 MHz and are negligibly small. 
The electric dipole in the disk was induced by the external electric field of the finite metasurface experiencing the local enhancement at the Fabry-Perot resonance. 
It was so because the polarized disk induced a dipole moment at the edge of the metasurface where charges were strongly accumulated on the ends of the wires. 
However, for our purposes this effect was not important. We confirmed that the frequency of MD mode was 270 MHz which we would explore further.

\subsection{WPT efficiency}
\begin{figure*}[t!]
\includegraphics[width=1\linewidth]{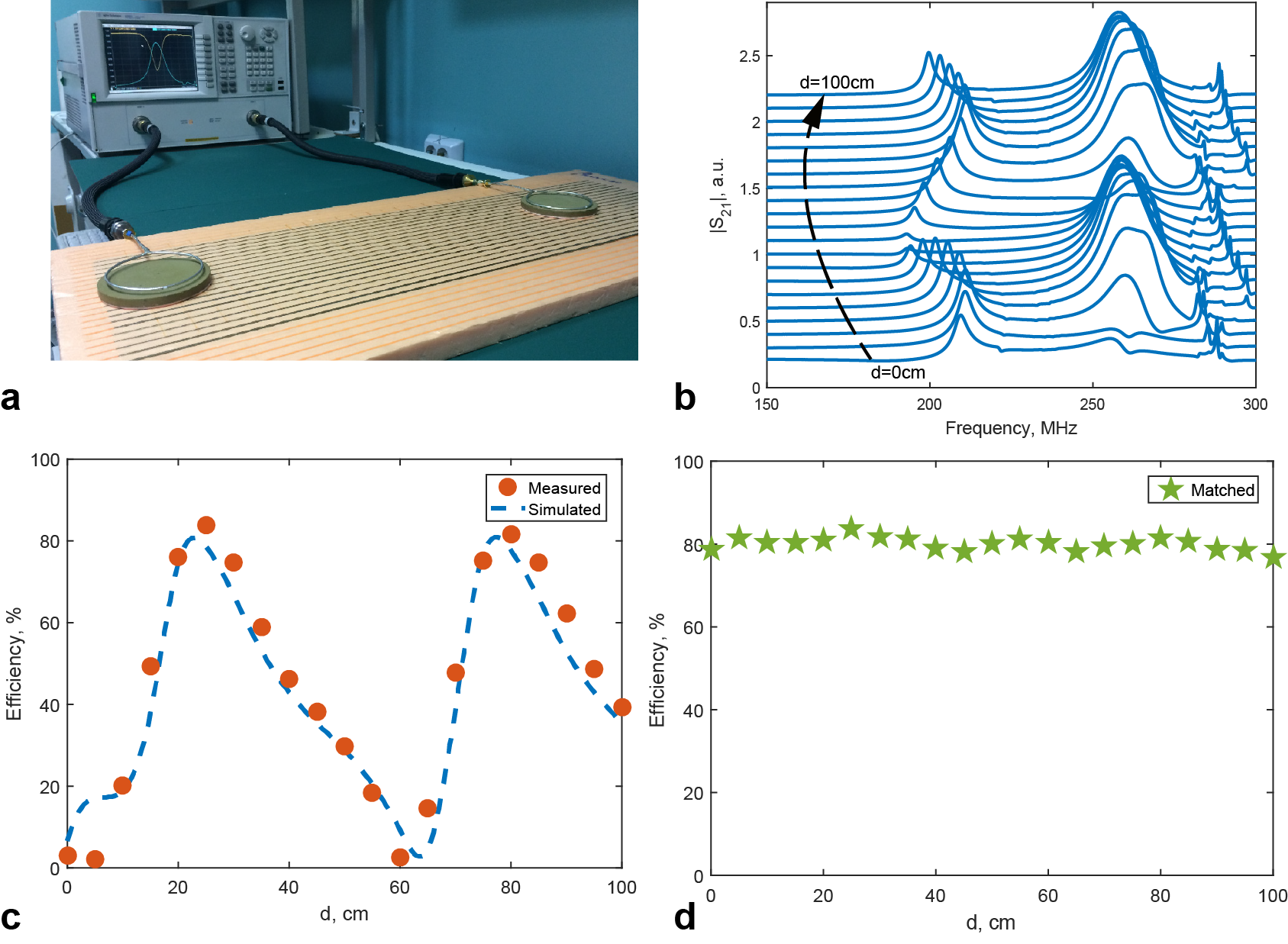}
\caption{(a) Photo of the experiment setup. (b) Measured transmission coefficient spectra for d ranging from 0 to 100 cm. (c) Comparison between measured and simulated efficiency (d)Measured WPT efficiency when both ports are matched. }
\label{fig6}
\end{figure*}

Now let's characterize the WPT system in terms of the efficiency depending on the distance $d$. We performed numerical simulations to investigate the S-parameter spectra as a function of $d$ ranging from 0 cm to 100 cm. The calculated WPT efficiency $\eta$ as a function of distance $d$ between the transmitter and the receiver is plotted in Fig. \ref{fig6}(c). It has a periodical tendency which can be explained by the presence of the standing wave features in the electric and magnetic field profiles. The maximal values of WPT efficiency as high as 80$\%$ were obtained at the distances of $d$=25 cm and $d$=80 cm where the magnetic field is maximal and the electric field is minimal at the center of the receiving disk. 
Domains where the efficiency was below 40\% cover less than one half of the interval of possible distances. Moreover, in these two intervals where $\eta$ was low the magnetic field kept sufficiently high for an efficient WPT. At d = 0 cm and d = 60 cm the ports in the transmitting and receiving loops were mismatched due to the over coupling through the metasurface, thereby leading to deep minima in the efficiency. The well-known method to partially cure these minima of efficiency is to introduce matching networks.~\cite{sophocles2003electromagnetic,stevens2015magnetoinductive, inagaki2014theory} 

Then we proceeded to the experimental studies. The first prototype of the smart table was fabricated (see Fig.~\ref{fig6}(a)) consisting of two identical disk resonators with $D = 84$ mm and $h = 6.6$ mm made from microwave ceramics based on $Ba,Sr,TiO_3$ solution doped with Mg.\cite{nenasheva2010low} To excite the resonator a non-resonant Faraday shielded loop with a diameter of $D_0 = 72$ mm was fabricated using a segment of a coaxial cable. The end of the Faraday loop was connected to the Agilent PNA E8362C Vector Network Analyzer (VNA). The Faraday loop was co-axially placed near the resonator at the distance of $s_1=1$ mm same as in the simulations. Both transmitter and receiver were placed at the distance $s_2=1$ mm above the metasurface which was fabricated as an array of copper wires with parameters as performed in the simulations.

The measured S-parameters as a function of frequency for different $d$ ranging from 0 to 100 cm are shown in Fig.\ref{fig6}(b). For any $d$ the maximum of $|S_{21}|$ occurred around 270 MHz, indicating that the MD mode we theoretically studied above was indeed excited in the system. By contrast, the first resonance frequency kept changing as $d$ increases which was a clear evidence of its nature as a dimensional resonance of the metasurface. Next, the WPT efficiency on the MD mode was obtained from the measured S-parameters and depicted as red dots in Fig.\ref{fig6}(c). The maximal efficiency of 83\% was obtained at $d$ = 25 cm. In this work, we employed a simple but effective method to match the whole system to reveal the maximal possible transfer efficiency. 
To match both ports we did not insert any lumped matching circuit. Instead, at each $d$ we mechanically tuned the distance $s_1$ and $s_2$ so that $|S_{21}|$ is maximal and both $|S_{11}|$ and $|S_{22}|$ are close to zero. This method is proved effective in Ref.~\onlinecite{song2016wireless1}.
The measurements of the S-parameters with this matching procedure was done for $d$ ranging from 0 to 100 cm with 5 cm step and the retrieved efficiency is depicted as green stars in Fig.\ref{fig6}(d). 
The WPT efficiency was almost stable at 80$\pm$3\% for all investigated distances $d$, which provided the possibility to place the receiver on the metasurface at any distances from the transmitter and it would be efficiently coupled to it. Moreover, the metasurface could have different dimensions, e.g. the metasurface length $L$ could be further increased to cover more area or decreased for customized needs. The only restriction of such WPT system was the misalignment between the transmitter and the receiver -- the power channel was stretched along the metasurface. But it could be solved by employing multiple transmitting resonators under the control of the power divider, as demonstrated in Fig.~\ref{fig1}(b).

\subsection{Safety Issue}

One of the target applications of such the WPT system is an ubiquitous wireless charging. Thus safety issues related to human exposure under electromagnetic fields (EMF) and specific absorption rate (SAR) must be taken into consideration. Here we considered SAR - a more rigorous safety metric for a measure of how much power is absorbed by biologic tissues.
If the proposed metasurface is used in a wireless charging system it can be embedded in a plastic or wooden desk. Thus we considered a typical scenario that a human arm was placed on top of the metasurface between the transmitter and the receiver. We investigated how much power can be absorbed by the human body by means of calculating the SAR.

\begin{figure*}[t]
\includegraphics[width=1\textwidth]{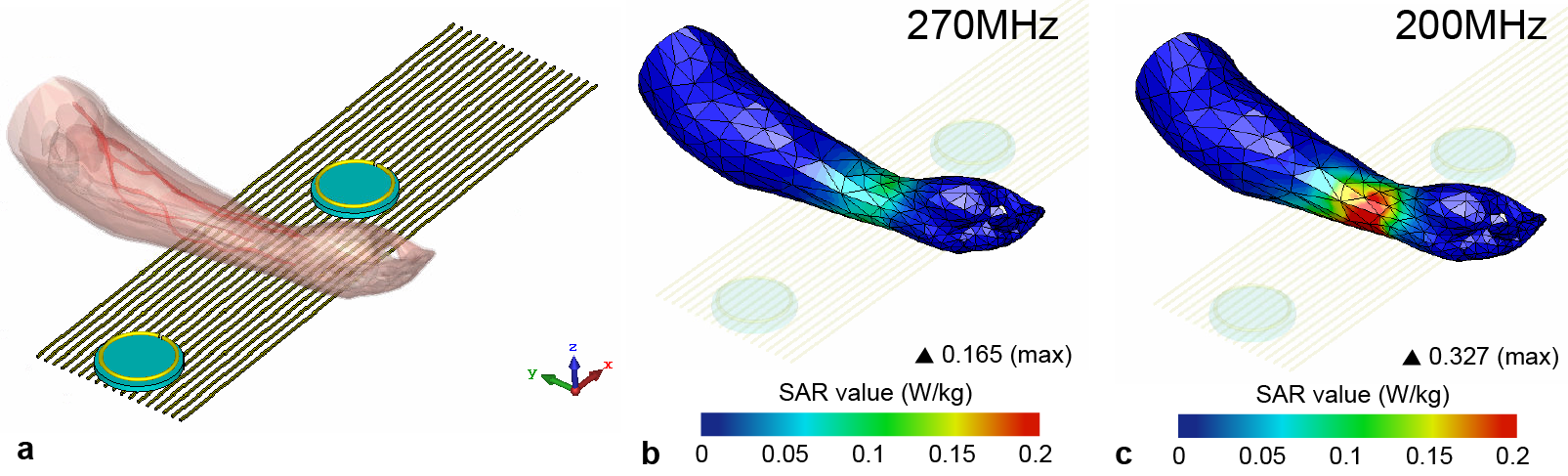}
\caption{(a) Simulation model of the WPT system with an arm located between the resonators 2 mm above the metasurface. Simulated SAR values at the frequency of (b) ED mode and (c) MD mode.}
\label{fig7}
\end{figure*}

To perform the SAR analysis we used CST Microwave Studio and a CAD model of the front part of a human arm (see Fig.\ref{fig7}(a)). The human arm model comprised the details of the main biological tissues of the arm (skin, fat, muscle, bone, blood, etc) which are characterized by corresponding electromagnetic properties.  The distance between the transmitter and the receiver was fixed to provide the maximum of the WPT efficiency without matching. The arm placed in middle was hanging at the height 2 mm from its bottom edge to the metasurface. Under a 0.5 W input power, the maximal SAR was 0.165 at the frequency of 270 MHz which corresponds to the MD mode (see Fig.\ref{fig7}(b)). For comparison we also simulated the SAR at the frequency of 200 MHz (ED mode) and found that it was 0.327 which is twice higher than the MD mode. There are no nonlinear effects in the WPT system, and maximum of SAR for different input power values could be obtained by scaling up these results. Thus, according to International Electrotechnical Commission standard~\cite{sar2017}, where the SAR limit is 2.0 W/kg averaged over 10 g of tissue absorbing the most signal, the maximal permitted input power of 6 W was allowed at the frequency of the MD mode and only 3 W for the ED mode.

\section{Conclusion}

In this paper, we discussed two typical scenarios where metasurfaces could be applied for wireless power transfer system. Especially, we proposed a first design of an cost-effective metasurface-based smart table as an intermediate between WPT transmitter and receiver. Both numerical and experimental studies verified the efficient and distance-independent WPT (efficiency higher than 80\%) for the case when the transmitting and receiving devices were both placed on the table. Though the metasurface operated like an effective transmission line our system was not similar to the well-known WPT systems with a capacitive coupling. The key difference was a uniquely efficient magnetic coupling of our transmitting/receiving device with proposed metasurface. This coupling was granted by the magnetic Mie resonance of a dielectric disk with colossal permittivity. This physical mechanism allowed us to successfully perform the resonant WPT along the metasurface. The only restriction of such WPT system was the misalignment between the transmitter and the receiver -- the power channel was stretched along the metasurface, which was mainly due to the extreme anisotropy of the metasurface and can be improved by an isotropic metasurface design. 

\section{Acknowledgment}

The authors are thankful for Georgiy Solomakha and Stanislav Glybovski for useful discussions. The calculations of multipole decomposition have been supported by the Russian Science Foundation (Project No. 17-72-10230). The numerical simulation and experimental investigation of the wireless power transfer system have been supported by the Russian Science Foundation (Project No. 17-79-20379). M.S. acknowledges the support from the China Scholarship Council (201508090124).

\bibliography{DiskOnWire_main}

\begin{thebibliography}{10}

\bibitem{glybovski2016Metasurface}
S.~B. Glybovski, S.~A. Tretyakov, P.~A. Belov, Y.~S. Kivshar, and C.~R.
  Simovski, ``Metasurfaces: From microwaves to visible,'' {\em Physics
  Reports}, vol.~634, pp.~1 -- 72, 2016.
\newblock Metasurfaces: From microwaves to visible.

\bibitem{chen2016Review}
H.-T. Chen, A.~J. Taylor, and N.~Yu, ``A review of metasurfaces: physics and
  applications,'' {\em Reports on Progress in Physics}, vol.~79, no.~7,
  p.~076401, 2016.

\bibitem{Ding2018Gradient}
F.~Ding, A.~Pors, and S.~I. Bozhevolnyi, ``Gradient metasurfaces: a review of
  fundamentals and applications,'' {\em Reports on Progress in Physics},
  vol.~81, no.~2, p.~026401, 2018.

\bibitem{Sievenpiper1999High}
D.~Sievenpiper, L.~Zhang, R.~F.~J. Broas, N.~G. Alexopolous, and
  E.~Yablonovitch, ``High-impedance electromagnetic surfaces with a forbidden
  frequency band,'' {\em IEEE Transactions on Microwave Theory and Techniques},
  vol.~47, pp.~2059--2074, Nov 1999.

\bibitem{slobozhanyuk2016enhancement}
A.~P. Slobozhanyuk, A.~N. Poddubny, A.~J. Raaijmakers, C.~A. van Den~Berg,
  A.~V. Kozachenko, I.~A. Dubrovina, I.~V. Melchakova, Y.~S. Kivshar, and P.~A.
  Belov, ``Enhancement of magnetic resonance imaging with metasurfaces,'' {\em
  Advanced materials}, vol.~28, no.~9, pp.~1832--1838, 2016.

\bibitem{yu2014flat}
N.~Yu and F.~Capasso, ``Flat optics with designer metasurfaces,'' {\em Nature
  materials}, vol.~13, no.~2, p.~139, 2014.

\bibitem{kildishev2013planar}
A.~V. Kildishev, A.~Boltasseva, and V.~M. Shalaev, ``Planar photonics with
  metasurfaces,'' {\em Science}, vol.~339, no.~6125, p.~1232009, 2013.

\bibitem{Khorasaninejad2016metalenses}
M.~Khorasaninejad, W.~T. Chen, R.~C. Devlin, J.~Oh, A.~Y. Zhu, and F.~Capasso,
  ``Metalenses at visible wavelengths: Diffraction-limited focusing and
  subwavelength resolution imaging,'' {\em Science}, vol.~352, no.~6290,
  pp.~1190--1194, 2016.

\bibitem{arbabi2018mems}
E.~Arbabi, A.~Arbabi, S.~M. Kamali, Y.~Horie, M.~Faraji-Dana, and A.~Faraon,
  ``Mems-tunable dielectric metasurface lens,'' {\em Nature communications},
  vol.~9, no.~1, p.~812, 2018.

\bibitem{ra2015thin}
Y.~Ra'Di, C.~Simovski, and S.~Tretyakov, ``Thin perfect absorbers for
  electromagnetic waves: theory, design, and realizations,'' {\em Physical
  Review Applied}, vol.~3, no.~3, p.~037001, 2015.

\bibitem{yu2011light}
N.~Yu, P.~Genevet, M.~A. Kats, F.~Aieta, J.-P. Tetienne, F.~Capasso, and
  Z.~Gaburro, ``Light propagation with phase discontinuities: generalized laws
  of reflection and refraction,'' {\em science}, p.~1210713, 2011.

\bibitem{zhao2012twisted}
Y.~Zhao, M.~Belkin, and A.~Al{\`u}, ``Twisted optical metamaterials for
  planarized ultrathin broadband circular polarizers,'' {\em Nature
  communications}, vol.~3, p.~870, 2012.

\bibitem{Hui2014Critical}
S.~Y.~R. Hui, W.~Zhong, and C.~K. Lee, ``A critical review of recent progress
  in mid-range wireless power transfer,'' {\em IEEE Transactions on Power
  Electronics}, vol.~29, pp.~4500--4511, Sept 2014.

\bibitem{lu2017review}
F.~Lu, H.~Zhang, and C.~Mi, ``A review on the recent development of capacitive
  wireless power transfer technology,'' {\em Energies}, vol.~10, no.~11,
  p.~1752, 2017.

\bibitem{song2017wireless}
M.~Song, P.~Belov, and P.~Kapitanova, ``Wireless power transfer inspired by the
  modern trends in electromagnetics,'' {\em Applied Physics Reviews}, vol.~4,
  no.~2, p.~021102, 2017.

\bibitem{assawaworrarit2017robust}
S.~Assawaworrarit, X.~Yu, and S.~Fan, ``Robust wireless power transfer using a
  nonlinear parity--time-symmetric circuit,'' {\em Nature}, vol.~546, no.~7658,
  p.~387, 2017.

\bibitem{krasnok2018coherently}
A.~Krasnok, D.~G. Baranov, A.~Generalov, S.~Li, and A.~Al{\`u}, ``Coherently
  enhanced wireless power transfer,'' {\em Physical review letters}, vol.~120,
  no.~14, p.~143901, 2018.

\bibitem{song2016wireless}
M.~Song, I.~Iorsh, P.~Kapitanova, E.~Nenasheva, and P.~Belov, ``Wireless power
  transfer based on magnetic quadrupole coupling in dielectric resonators,''
  {\em Applied Physics Letters}, vol.~108, no.~2, p.~023902, 2016.

\bibitem{costanzo2014electromagnetic}
A.~Costanzo, M.~Dionigi, D.~Masotti, M.~Mongiardo, G.~Monti, L.~Tarricone, and
  R.~Sorrentino, ``Electromagnetic energy harvesting and wireless power
  transmission: A unified approach,'' {\em Proceedings of the IEEE}, vol.~102,
  no.~11, pp.~1692--1711, 2014.

\bibitem{kurs2007wireless}
A.~Kurs, A.~Karalis, R.~Moffatt, J.~D. Joannopoulos, P.~Fisher, and
  M.~Solja{\v{c}}i{\'c}, ``Wireless power transfer via strongly coupled
  magnetic resonances,'' {\em science}, vol.~317, no.~5834, pp.~83--86, 2007.

\bibitem{karalis2008efficient}
A.~Karalis, J.~D. Joannopoulos, and M.~Solja{\v{c}}i{\'c}, ``Efficient wireless
  non-radiative mid-range energy transfer,'' {\em Annals of physics}, vol.~323,
  no.~1, pp.~34--48, 2008.

\bibitem{urzhumov2011metamaterial}
Y.~Urzhumov and D.~R. Smith, ``Metamaterial-enhanced coupling between magnetic
  dipoles for efficient wireless power transfer,'' {\em Physical Review B},
  vol.~83, no.~20, p.~205114, 2011.

\bibitem{chen2012extremely}
W.-C. Chen, C.~M. Bingham, K.~M. Mak, N.~W. Caira, and W.~J. Padilla,
  ``Extremely subwavelength planar magnetic metamaterials,'' {\em Physical
  Review B}, vol.~85, no.~20, p.~201104, 2012.

\bibitem{dong2017enhanced}
Z.~Dong, F.~Yang, and J.~S. Ho, ``Enhanced electromagnetic energy harvesting
  with subwavelength chiral structures,'' {\em Physical Review Applied},
  vol.~8, no.~4, p.~044026, 2017.

\bibitem{wang2011experiments}
B.~Wang, K.~H. Teo, T.~Nishino, W.~Yerazunis, J.~Barnwell, and J.~Zhang,
  ``Experiments on wireless power transfer with metamaterials,'' {\em Applied
  Physics Letters}, vol.~98, no.~25, p.~254101, 2011.

\bibitem{ho2015planar}
J.~S. Ho, B.~Qiu, Y.~Tanabe, A.~J. Yeh, S.~Fan, and A.~S. Poon, ``Planar
  immersion lens with metasurfaces,'' {\em Physical Review B}, vol.~91, no.~12,
  p.~125145, 2015.

\bibitem{simovski2012wire}
C.~R. Simovski, P.~A. Belov, A.~V. Atrashchenko, and Y.~S. Kivshar, ``Wire
  metamaterials: physics and applications,'' {\em Advanced Materials}, vol.~24,
  no.~31, pp.~4229--4248, 2012.

\bibitem{rahman2009importance}
A.~Rahman, P.~A. Belov, M.~G. Silveirinha, C.~R. Simovski, Y.~Hao, and
  C.~Parini, ``The importance of fabry--perot resonance and the role of
  shielding in subwavelength imaging performance of multiwire endoscopes,''
  {\em Applied Physics Letters}, vol.~94, no.~3, p.~031104, 2009.

\bibitem{rahman2010periscope}
A.~Rahman, P.~A. Belov, Y.~Hao, and C.~Parini, ``Periscope-like endoscope for
  transmission of a near field in the infrared range,'' {\em Optics letters},
  vol.~35, no.~2, pp.~142--144, 2010.

\bibitem{belov2010experimental}
P.~A. Belov, G.~K. Palikaras, Y.~Zhao, A.~Rahman, C.~R. Simovski, Y.~Hao, and
  C.~Parini, ``Experimental demonstration of multiwire endoscopes capable of
  manipulating near-fields with subwavelength resolution,'' {\em Applied
  Physics Letters}, vol.~97, no.~19, p.~191905, 2010.

\bibitem{radu2009toward}
X.~Radu, D.~Garray, and C.~Craeye, ``Toward a wire medium endoscope for mri
  imaging,'' {\em Metamaterials}, vol.~3, no.~2, pp.~90--99, 2009.

\bibitem{tuniz2013metamaterial}
A.~Tuniz, K.~J. Kaltenecker, B.~M. Fischer, M.~Walther, S.~C. Fleming,
  A.~Argyros, and B.~T. Kuhlmey, ``Metamaterial fibres for subdiffraction
  imaging and focusing at terahertz frequencies over optically long
  distances,'' {\em Nature communications}, vol.~4, p.~2706, 2013.

\bibitem{slobozhanyuk2014endoscope}
A.~Slobozhanyuk, I.~Melchakova, A.~Kozachenko, D.~Filonov, C.~Simovski, and
  P.~Belov, ``An endoscope based on extremely anisotropic metamaterials for
  applications in magnetic resonance imaging,'' {\em Journal of Communications
  Technology and Electronics}, vol.~59, no.~6, pp.~562--570, 2014.

\bibitem{song2016wireless1}
M.~Song, P.~Belov, and P.~Kapitanova, ``Wireless power transfer based on
  dielectric resonators with colossal permittivity,'' {\em Applied Physics
  Letters}, vol.~109, no.~22, p.~223902, 2016.

\bibitem{evlyukhin2016optical}
A.~B. Evlyukhin, T.~Fischer, C.~Reinhardt, and B.~N. Chichkov, ``Optical
  theorem and multipole scattering of light by arbitrarily shaped
  nanoparticles,'' {\em Physical Review B}, vol.~94, no.~20, p.~205434, 2016.

\bibitem{terekhov2017multipolar}
P.~D. Terekhov, K.~V. Baryshnikova, Y.~A. Artemyev, A.~Karabchevsky, A.~S.
  Shalin, and A.~B. Evlyukhin, ``Multipolar response of nonspherical silicon
  nanoparticles in the visible and near-infrared spectral ranges,'' {\em
  Physical Review B}, vol.~96, no.~3, p.~035443, 2017.

\bibitem{sophocles2003electromagnetic}
J.~O. Sophocles, {\em Electromagnetic waves and antennas}.
\newblock 2003.

\bibitem{stevens2015magnetoinductive}
C.~J. Stevens, ``Magnetoinductive waves and wireless power transfer,'' {\em
  IEEE Transactions on Power Electronics}, vol.~30, no.~11, pp.~6182--6190,
  2015.

\bibitem{inagaki2014theory}
N.~Inagaki, ``Theory of image impedance matching for inductively coupled power
  transfer systems,'' {\em IEEE Transactions on Microwave Theory and
  Techniques}, vol.~62, no.~4, pp.~901--908, 2014.

\bibitem{nenasheva2010low}
E.~Nenasheva, N.~Kartenko, I.~Gaidamaka, O.~Trubitsyna, S.~Redozubov, A.~Dedyk,
  and A.~Kanareykin, ``Low loss microwave ferroelectric ceramics for high power
  tunable devices,'' {\em Journal of the European Ceramic Society}, vol.~30,
  no.~2, pp.~395--400, 2010.

\bibitem{sar2017}
``Determination of rf field strength, power density and sar in the vicinity of
  radiocommunication base stations for the purpose of evaluating human
  exposure,'' tech. rep., IEC, 2017.

\end{thebibliography}
\bibliographystyle{unsrt}

\end{document}